\documentclass[aps,pre,twocolumn]{revtex4}
\usepackage{epsfig}
\begin{document}
\def\I.#1{\it #1}
\def\B.#1{{\bf #1}}
\def\C.#1{{\cal  #1}}
\newcommand{\beq}{\begin{equation}}
\newcommand{\eeq}{\end{equation}}
\newcommand{\bea}{\begin{eqnarray}}
\newcommand{\eea}{\end{eqnarray}}
\title{Bi-Laplacian Growth Patterns in Disordered Media}
   \author { Anders Levermann and Itamar Procaccia}
   \affiliation{Department of~~Chemical Physics, The Weizmann Institute of
   Science, Rehovot 76100, Israel}
%
%
\begin{abstract}
Experiments in quasi 2-dimensional geometry (Hele Shaw cells) in which
a fluid is injected into a visco-elastic medium (foam, clay or associating-
polymers) show patterns akin to fracture in brittle materials, very
different from standard Laplacian growth patterns of viscous fingering. An
analytic theory
is lacking since a pre-requisite to describing the fracture of elastic material is 
the solution of the bi-Laplace rather than
the Laplace equation. In this Letter we close this gap, offering a theory
of bi-Laplacian growth patterns based on the method of iterated conformal maps.
\end{abstract}
 \maketitle
Pattern formation for two-phase flow instabilities has been intensely studied,
both experimentally and theoretically, for the displacement of a viscous
fluid from between parallel plates or from a porous medium \cite{86DKLST}. In these 
cases the velocity field $\B.v(r)$ is well described by Darcy's law 
$\B.v(\B.r)\propto \B.\nabla P(\B.r)$, where $P(\B.r)$ is the pressure.
For incompressible fluids $\B.\nabla \cdot \B.v=0$, leading to the
Laplace equation for the pressure, $\nabla^2 P(\B.r)=0$, with appropriate
boundary condition on the boundary of the growing pattern and at ``infinity".
The theory for such ``Laplacian Growth" patterns in 2-dimensions (i.e.
$\B.r=(x,y)$) naturally 
focuses on analytic functions (or their conformal inverse), simply because
the Cauchy-Riemann conditions imply that the general solution of the
Laplace equation is given by the real part of an analytic function,
$P=\Re \{F(z)\}$, where $F(z)$ is the unique analytic function that
satisfies the boundary conditions, and $z=x+iy$ \cite{58ST,84SB}.

\begin{figure}
\epsfxsize=8cm 
\epsfbox{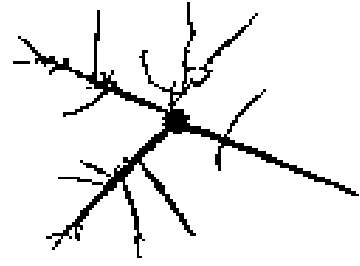}
\caption{Typical pattern when water is injected into a radial Hele-Shaw cell filled with a
solution of associated polymers \cite{93ZM}}
\label{Fig.1}
\end{figure}

Sporadically, over the last decade, there appeared experimental studies in 
which a low viscosity fluid displaces not a more viscous fluid, but rather
a medium which is visco-elastic, like foam \cite{99LRC}, clay \cite{91LLDD}, or a solution of
associating polymers \cite{93ZM}.  Elastic media are expected to be invaded by fracture, rather
than a displacement, and indeed the growth patterns reported in the experiments
had features akin to fracture patterns in brittle materials, see Fig. \ref{Fig.1}. 
Detailed comparisons with theory were lacking however, since an appropriate 
analytic theory did not exist.
As is well known, (and see below for details), in fracture the relevant
equation to solve is the bi-Laplace equation $\nabla^2\nabla^2 \chi=0$ with
appropriate boundary conditions \cite{86LL}. The general solution is no longer the
real part of an analytic function, but rather 
\begin{equation}
\chi(z,\bar z) = \Re \left[ \overline{z} 
\phi(z) + \tilde \psi(z) \right] \ ,\label{solbi}
\end{equation} where
$\phi(z)$ and $\tilde \psi(z)$ are a pair of analytic functions. Thus conformal techniques
are not trivially applicable, and until recently there was no appropriate 
theoretical method to solve such equations with boundary conditions on
an arbitrary ramified boundary. Even numerical simulations were limited to 
lattice discretizations \cite{93WC}, even though
lattice anisotropy is a relevant perturbation changing the universality class
of the growing patterns. Recent progress in the context of quasi-static
fracture \cite{02BHLP,02BLP} allows us to offer below an appropriate model for bi-Laplacian
Growth patterns.

To set up the model imagine a 2-dimensional elastic medium with a hole of an
arbitrary shape, whose boundary $z(s)$ is parametrized by the arc-length
variable $s$. Into this hole
one pushes quasistatically a fluid of pressure $P$. In equilibrium with this
pressure the elastic medium will suffer a displacement field $u(\B.r)$. The strain 
tensor $\epsilon_{jk}$ which results is
\begin{equation}
\epsilon_{jk}\equiv\frac{1}{2} \left( \partial_j u_k + \partial_k u_j \right) \ . \label{strain}
\end{equation}
In linear elasticity theory \cite{86LL} the stress tensor is related to the strain tensor by
\begin{equation}
\sigma_{jk} = \frac{E}{1+\sigma} \left( u_{jk} + \delta_{jk} \; \frac{\sigma}{1-2\sigma}
\sum_l u_{ll} \right) \ , \label{stress}
\end{equation}
where $E$ and $\sigma$ are material parameters. Equilibrium inside the elastic medium
requires that 
\begin{equation}
\sum_k \partial_{k} \; \sigma_{jk} = 0 \quad \mbox{ for all } j \label{equi1}
\end{equation}
The general solution of these equations in 2-dimensional is given by 
\begin{equation}
\sigma_{xx}  =  \partial_y^2 \chi \ , \quad
\sigma_{yy}  =  \partial_x^2 \chi \ , \quad
\sigma_{xy}  =  -\partial_{xy} \chi  \ , \label{sigmachi}
\end{equation}
where the so called Airy potential $\chi$ fulfills the biharmonic equation
\begin{equation}
\nabla^2 \nabla^2 \chi = 0 \ .
\end{equation}
The solution is represented, as said, by Eq. (\ref{solbi}). In order to develop
the growth pattern we need to compute the tangent component of the stress tensor
at the boundary of the pattern,
since cracking proceeds only if this component exceeds a threshold $\sigma_c$.
To define this component and to state the conditions on the boundary of the growth pattern we use
the local tangent and normal directions. With $\alpha$ 
being the angle 
between the tangent and the $x$-axis at $z(s)$, we define derivatives with respect to the
tangent and normal directions according to 
\begin{eqnarray}
\partial_t & = & \cos(\alpha) \; \partial_x + \sin(\alpha) \; \partial_y \nonumber \\
\partial_n & = & \cos(\alpha) \; \partial_y - \sin(\alpha) \; \partial_x \ . \label{tannor}
\end{eqnarray}
The pressure $P$ now must be balanced by the normal component of the stress:
\begin{equation}
\partial_{tt} \chi = \sigma_{nn} = - P = const. \quad \mbox{ on the crack} \ . \label{tantan}
\end{equation}
Since in equilibrium no fluid slips along the boundary,
\beq
-\partial_{tn} \chi = \sigma_{tn} = \sigma_{nt} = 0 \quad \mbox{ on the crack} \ . \label{noslip}
\eeq
The normal component $\partial_{nn} \chi = \sigma_{tt}$ is not determined from the
boundary conditions, but is {\em computed} by solving for $\chi(z,\bar z)$.  Using the
fact that $4\partial^2\chi/\partial z\partial\bar z=\sigma_{xx}+\sigma_{yy}=
\sigma_{tt}+\sigma_{nn}$ we can immediately read from Eq. (\ref{solbi}),
\begin{equation}
\sigma_{tt}(z)=P+4\Re[\phi'(z)] \ , \quad \text{at the boundary} \ . \label{solution}
\end{equation}
Once we have computed the tangent component of the stress, we may advance
the crack if $\Delta \sigma\equiv \sigma_{tt}(z)-\sigma_c> 0$, at a speed
which is (on the average) proportional to $\Delta \sigma$ \cite{90Ker,90HR}.

Thus to compute the tangent stress and advance the crack we only need to determine
the function $\phi(z)$. The boundary conditions (\ref{tantan}) and (\ref{noslip}) 
are expressed in terms of $\phi(z)$ and $\tilde \psi(z)$ by using (\ref{tannor})
in (\ref{tantan}) and (\ref{noslip}), to derive
$\partial_t \partial_{\bar z} \chi =-P [\cos(\alpha)+i\sin(\alpha)]=-P\partial_t z(s)$,
where we identify $\partial_t z(s)$ as the unit vector tangent to the boundary.
Rewriting this condition as $\partial_t [\partial_{\bar z} \chi +Pz]=0$
we obtain the boundary condition on the interface \cite{52Mus}
\begin{equation}
\phi(z(s)) + z(s) \;
\overline{\phi^\prime(z(s))} + \overline{\psi(z(s))} = - P \; z(s) + K
\end{equation}
where $\psi(z)\equiv \tilde \psi'(z)$ and $K$ is a constant that can be chosen
zero with impunity.

The boundary conditions at infinity are obvious, since all stress components have to
vanish as $z \rightarrow \infty$:
\begin{equation}
\partial_{z\bar z} \chi(z,\bar z)\rightarrow 0, ~\partial_{zz} \chi(z,\bar z) \rightarrow 0
\quad \mbox{ as } z \rightarrow \infty
 \ . \label{bcinfty}
\end{equation}
In light of these conditions $\phi(z)$ must have the form
$\phi(z) = i \beta_1 z + \sum_{j=0}^{\infty} u_{-j} z^{-j}$ 
with $\beta_1$ real. The solution of the stress field is invariant under the
transformation $ \phi \rightarrow \phi + i \; A \; z + B$ 
with $A$ real and $B$ a complex constant. We can use this freedom to get rid of
$\beta_1$ and $u_0$, and write $\phi$ in the form
\begin{equation}
\phi(z) = \sum_{j=1}^{\infty} u_{-j} z^{-j} \ . \label{phi}
\end{equation}
Similarly from (\ref{bcinfty}) it follows that $\psi$ has the form
\begin{equation}
\psi(z) = \sum_{j=1}^{\infty} v_{-j} z^{-j} \ . \label{psi}
\end{equation}

To proceed invoke a conformal map $z = \Phi^{(n)}(\omega )$ that maps the exterior of 
the unit circle in the mathematical plane
$\omega$ to the exterior of the crack in the physical plane $z$, after $n$ growth steps. The
conformal map is univalent by construction, with a 
Laurent expansion
\begin{equation}
\Phi^{(n)}(\omega ) = F_1^{(n)}\omega + F_0^{(n)} 
+F_{-1}^{(n)}/\omega+F_{-2}^{(n)}/\omega^2+\cdots \ . \label{Laurent}
\end{equation}
and $\Phi^{(0)}(\omega )=\omega$. The arclength position
$s$ in the physical domain is mapped by the inverse of $\Phi^{(n)}$ onto
a position on the unit circle $\epsilon=\exp(i\theta)$.
We will be able to compute the stress tensor on the boundary of 
the crack in the physical domain by 
performing the calculation on the unit circle. In other words we
will compute $\sigma_{tt}(\theta)$ on the unit circle in the mathematical plane. 

We perform the
calculation iteratively, taking the stress as known for the crack after $n-1$ fracture events.
In order
to implement the $n$th cracking event with average velocity 
proportional to $\Delta \sigma$,
we should choose potential positions on the interface
more often when $\Delta\sigma(\theta)$ is larger. We construct
a probability density $P(\theta)$ on the unit circle $e^{i\theta}$
which satisfies
\begin{equation}
P(\theta) = \frac{|\Phi^{'(n-1)}(e^{i \theta} )|\Delta\sigma(\theta)
\Theta(\Delta\sigma(\theta))}{\int_0^{2\pi}|\Phi^{'(n-1)}(e^{i \tilde\theta} )|\Delta\sigma(\tilde\theta)
\Theta(\Delta\sigma(\tilde\theta))d\tilde\theta}
\ , \label{weight}
\end{equation} 
where $\Theta(\Delta\sigma(\tilde\theta))$ is the Heaviside function, 
and $|\Phi^{'(n-1)}(e^{i \theta} )|$ is simply the Jacobian of the 
transformation from mathematical to physical plane. The next
growth position, $\theta_n$ in the mathematical plane, is chosen randomly with
respect to the probability $P(\theta) d\theta$. 
At the chosen position on the crack, i.e. $z= \Phi^{(n-1)}(e^{i\theta_n})$, we want to advance
the crack with a given step of fixed length $\sqrt{\lambda_0}$. We achieve growth with an auxiliary
conformal map
$\phi_{\lambda_n,\theta_n}(\omega )$ that maps the unit circle to a unit circle with a 
semi-circular bump
of area $\lambda_n$ centered at $e^{i\theta_n}$ \cite{98HL,99DHOPSS}. 
To ensure a fixed size step in the physical domain we choose
\begin{equation}
   \lambda_{n} = \frac{\lambda_0}{|{\Phi^{(n-1)}}' (e^{i \theta_n})|^2} \ .
   \label{lambdan}
\end{equation}
Finally the updated conformal map $\Phi^{(n)}$ is obtained as
\begin{equation}
\label{conformal}
\Phi^{(n)}(\omega ) = \Phi^{(n-1)}(\phi_{\lambda_n,\theta_n}(\omega )) \ . \label{iter}
\end{equation} 

The recursive dynamics can be represented as iterations
of the map $\phi_{\lambda_{n},\theta_{n}}(w)$,
\begin{equation}
   \Phi^{(n)}(w) =
\phi_{\lambda_1,\theta_{1}}\circ\phi_{\lambda_2,\theta_{2}}\circ\dots\circ
\phi_{\lambda_n,\theta_{n}}(\omega)\ . \label{comp}
\end{equation} 
Every given fracture pattern is determined completely by the random itinerary
$\{\theta_i\}_{i=1}^n$.

We can now represent the boundary conditions (\ref{solution}) in terms
of $z(s)= \Phi^{(n)}(\epsilon )$:
\begin{equation}
\phi(\Phi^{(n)}(\epsilon )) + \Phi(\epsilon) \;
\frac{\overline{\phi^\prime(\Phi^{(n)}(\epsilon ))}}{\overline{\Phi^{(n)\prime}(\epsilon)}} +
\overline{\psi(\Phi^{(n)}(\epsilon ))} = - P  \Phi^{(n)}(\epsilon )  \ . \label{bceps}
\end{equation}
to solve this equation we introduce a power expansion for the ratio
\begin{eqnarray}
\frac{\Phi^{(n)}(\epsilon)}{\overline{\Phi^{(n)\prime}(\epsilon)}}  = \sum_{j=-\infty}^{\infty}
b_j \; \epsilon^j \ . \label{Fourier}
\end{eqnarray}
Note that this expansion contains both positive and negative powers of $\epsilon$,
whereas Eqs. (\ref{phi}) and (\ref{psi}) contained only negative powers.
Nevertheless upon substituting all the power expansions into Eq. (\ref{bceps}) one
finds that the determination of the function $\phi(z)$ only requires the negative
powers of $\epsilon$ \cite{02BLP}, with the coefficients satisfying the system of 
equations
\begin{equation}
u_{-l} - \sum_{j=1}^{\infty} \; j \; \overline{u}_{-j} \; b_{-(j+l+1)} = - p \; F_{-l} \ .
\label{system}
\end{equation}
After separating real from imaginary parts one finds an infinite system of linear
equations. In practice we truncate at $j_{\rm max}=100$ and test for convergence by increasing
the order. Note that the highest resolved $u_{j_{\rm max}}$ requires computing the Fourier
series (\ref{Fourier}) to order $2j_{\rm max}+1$.

\begin{figure}
\epsfxsize=8cm \epsfbox{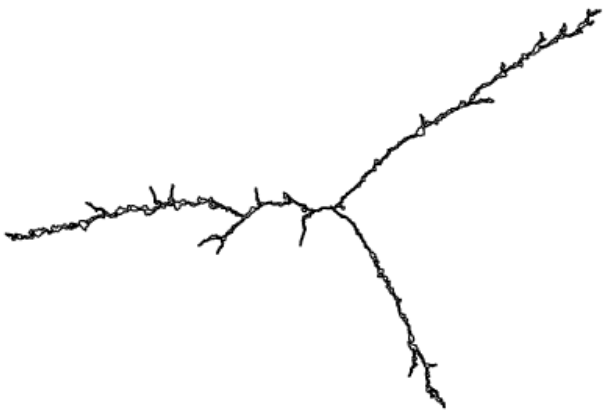}
\caption{Typical pattern resulting from a growth of discrete fracture events
occuring with the probability $P \approx \Theta(\Delta \sigma) \Delta \sigma$.}
\label{Fig.2}
\end{figure}

Implementing this procedure with $\Phi^{(0)}(\omega)=\omega$, and choosing 
$\lambda_0=1$, we generate a typical fracture pattern as seen in Fig. \ref{Fig.2}.
What is seen is the map $\Phi^{(4500)}(\epsilon)$ which is topologically a circle.

While we can guarantee that the pattern seen is indeed an exact bi-Laplacian
pattern under the growth rules adopted here, we cannot guarantee that it
is identical to any of the experimental patterns reported in \cite{99LRC,91LLDD,93ZM}. 
This stems from a few reasons. First, in many experiments there is a mixture of
viscous and elastic phenomena, to the point that there are examples of a continuum
of growth patterns depending on the relative importance of the two \cite{91LLDD,93ZM}. In our theory 
we solved the bi-Laplacian equation after each growth step; this is relevant in the
purely elastic limit. Secondly, and not less importantly, we advanced the pattern
where allowed ($\Delta\sigma>0$) at a velocity that is proportional (on the average)
to $\Delta\sigma$. While this is accepted by a number of authors as a reasonable guess
for the rate of growth of a fracture pattern in the quasi-static limit, it is by
no means derived from first principles or universally accepted. Needless to say,
in our procedure we can adopt any other velocity law without much ado, simply by
changing the probability distribution  (\ref{weight}). We caution the reader that 
one does not expect the patterns to be independent of the velocity law \cite{02BLP}. Thus 
a more complete theory of bi-Laplacian patterns calls for further collaboration
between experiments and theory to zero in on a plausible velocity law. Before doing
so there is a limited relevance to studying carefully the geometric properties of the
patterns obtained with this velocity law or another.
\begin{figure}
\epsfxsize=8cm 
\epsfbox{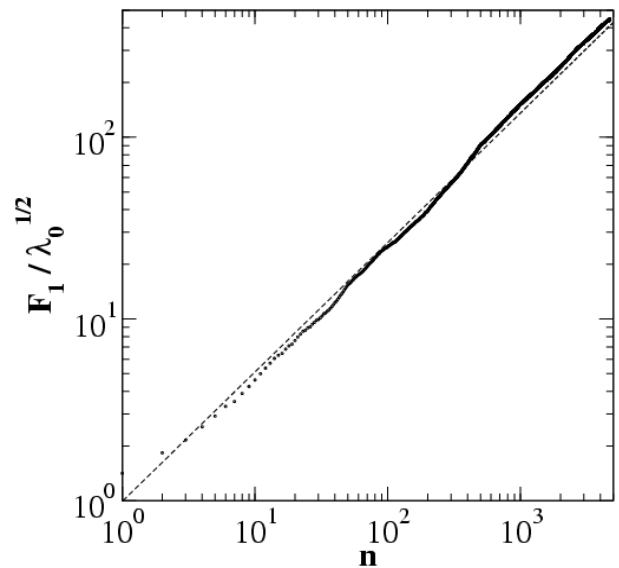}
\caption{$F_1^{(n)}/\sqrt{\lambda_0}$ of the pattern in Fig. \ref{Fig.2} vs. $n$ in a double
logarithmic plot. From the slope of the least squares fit we estimate the
dimension $D=1.4\pm 0.1$.}
\label{Fig.3}
\end{figure}
Notwithstanding these remarks, we stress that the present theory offers a 
very convenient tool for assessing the fractal dimension $D$ of the growing
patterns. Having a univalent conformal map as in (\ref{Laurent}), one can
invoke the rigorous ``1/4" theorem. This theorem states that if $R_n$ is the
radius of the minimal circle that contains the pattern after $n$ fracture events,
then 
\begin{equation}
F_1^{(n)} \ge \frac{1}{4} R_n \ . \label{1fourth}
\end{equation}
Accordingly one expects that for large $n$ the first Laurent coefficient
satisfies
\begin{equation}
F_1^{(n)} \approx \sqrt{\lambda_0} ~n^{1/D} \ . \label{F1scales}
\end{equation}
One advantage of the present approach is that the first Laurent coefficient 
$F_1^{(n)}$ is known exactly as
\begin{equation}
F_1^{(n)} = \Pi_{k=1}^n \sqrt{1+\lambda_k},
\end{equation}
which is computable to machine precision. In Fig. \ref{Fig.3} we
present, in double logarithmic plot, $F_1^{(n)}/\sqrt{\lambda_0}$ vs. $n$.
Reading the slope of the least-squares fit yields a dimension $D=1.4 \pm 0.1$.

The method of iterated conformal maps offers convergent calculations
of fractal and multifractal properties of the growth patterns.
If required, one can use the formalism to
obtain highly accurate values of the fractal dimension, cf. \cite{00DLP}.
In addition, if the properties of the growth probabilities are of
interest from the multifractal point of view, there are available
methods to compute these in a convergent scheme \cite{01DJLMP} that is not available
in direct numerical simulations. However, such refinements
would be justified only after future work to solidify further the 
relation between theory and experiment.

\acknowledgments
We thank Jonathan Tal for his expert help with the SV1
supercomputer at the Weizmann Institute.
This work has been supported in part by the Minerva Foundation,
Munich, Germany, the
Petroleum Research Fund, The  European Commission under the
TMR program and the Naftali and Anna
Backenroth-Bronicki Fund for Research in Chaos and Complexity.


\begin{thebibliography}{99}
\bibitem{86DKLST}
D. Bensimon, L. P. Kadanoff, S. Liang, B. I. Shraiman and C. Tang, 
Rev. Mod. Phys. {\bf 58}, 977 (1986).

 \bibitem{58ST}
   P.G. Saffman and G.I. Taylor, Proc. Roy. Soc. London Series A,
   {\bf 245},312 (1958).

 \bibitem{84SB}
   B. Shraiman and D. Bensimon,
   Phys.Rev. A{\bf 30}, 2840 (1984);
   S.D. Howison, J. Fluid Mech. {\bf 167}, 439 (1986).


\bibitem{99LRC}
A. Lindner, S. Rica and Y. Couder, Recontre du non-Lineaire,
p. 68 (1999).

\bibitem{91LLDD}
E. Lemaire, P. Levitz, G. Daccord and H. van Damme, Phys. Rev. Lett., {\bf 67}, 2009 (1991).

\bibitem{93ZM}
H. Zhao and J. V. Maher, Phys. Rev. E {\bf 47}, 4278 (1993).

\bibitem{86LL}
L.D. Landau and E.M. Lifshitz, {\em Theory of Elasticity}, 3rd ed. (Pergamon, London, 1986).

\bibitem{93WC}
W. Wang and W. Canessa, Phys. Rev. E {\bf 47}, 1243 (1993).

\bibitem{02BHLP}
F. Barra, H. G. E. Hentschel, A. Levermann and I. Procaccia, Phys. Rev.
       E., {\bf 65}, 045101 (2002) 

\bibitem{02BLP}
F. Barra, A. Levermann and I. Procaccia, ``Quasi-Static Brittle Fractures in Inhomogeneous Media and
Iterated Conformal Maps: Modes I, II and III", Phys. Rev. E,
       submitted. Also: cond-mat/0205132 

\bibitem{90Ker}
See for example J. Kert\'esz in \cite{90HR}.

\bibitem{90HR}
H.J. Herrmann and S. Roux, {\em Statistical Models for the Fracture
of Disordered Media}, (North Holland, Amsterdam, 1990), and references
therein.

\bibitem{52Mus}
N.I. Muskhelishvili, {\em Some Basic Problems in the Mathematical Theory of
Elasticity}, (Noordhoff, Groningen, 1952).

\bibitem{98HL}
M.B. Hastings and L.S. Levitov, Physica D {\bf 116},
244 (1998).

\bibitem{99DHOPSS} 
B. Davidovitch, H.G.E. Hentschel, Z. Olami, I.Procaccia, L.M. Sander, and E. Somfai,
   Phys. Rev. E, {\bf 59} 1368 (1999).


\bibitem{00DLP}
B. Davidovitch, A. Levermann, I. Procaccia, Phys. Rev. E, {\bf 62} R5919-R5922(2000).

\bibitem{01DJLMP}
B. Davidovitch, M. H. Jensen, A. Levermann, 
J. Mathiesen and I. Procaccia,  Phys.
Rev. Lett. {\bf 87}, 164101 (2001). 




\end{thebibliography}
\end{document}